\documentclass[12pt,prl,preprint,showpacs,showkeys]{revtex4}
\usepackage{amsfonts}
\usepackage{graphicx}
\usepackage{dcolumn}
\usepackage[version=3]{mhchem}
\usepackage{textcomp}
\usepackage{bm}
\include{biblio}

\begin{document}

\title{$Ab$ $initio$ study of the adsorption and dissociation of nitrogen molecule on Fe(111) surface}

\author{Myong-Song Ryang, Nam-Hyok Kim, Song-Jin Im}
\affiliation{Department of Physics, Kim Il Sung University, Daesong District, Pyongyang, DPR Korea}

\begin{abstract}
The adsorption and dissociation of nitrogen molecule on \ce{Fe}(111) surface is studied  by density functional theory calculations. The simulation results show that the molecule needs to acquire parallel orientation with respect to the surface for the adsorption and dissociation. In addition, \ce{Fe}(111) surface is more active in dissociating \ce{N2} than \ce{Fe}(100) surface due to its morphology. The interaction between antibonding molecular orbitals of \ce{N2} and partially filled $3d$ orbitals of \ce{Fe} atoms on the surface may be the key to the molecular dissociation of \ce{N2}. To break down the triple bond of \ce{N2}, the electron density on the surface needs to be partially transferred to the molecule to fill the antibonding molecular orbitals of the nitrogen molecule. The present result may provide some insights on the dissociation mechanism of molecules over transition metal surfaces.
\end{abstract}


\maketitle

\section{1.	Introduction}
It is practically important for the development of heterogeneous catalysts to study the adsorption and dissociation process of various gas molecules on transition metal surfaces on a microscopic scale. Furthermore, it is important in the basic research field to verify the change in the electronic structure during the adsorption and dissociation of gas molecules. The adsorption and dissociation of gas phase molecules on various surfaces has been intensively studied\cite{2, 4, 6,7, 11, 12, 13, 14, 15}. Dahl $et~ al$\cite{12} studied the molecular dissociation of nitrogen on \ce{Ru}(0001) surface and showed that it lowered the energy barrier for the molecular dissociation of nitrogen significantly. In particular, the study on the catalytic activity of various transition metal surfaces in dissociating nitrogen molecules shows that, along with \ce{Ru}, \ce{Fe} is a promising candidate as a catalyst for the dissociation of nitrogen molecule \cite{1,6}.
The filling of the antibonding molecular orbitals of the nitrogen molecule is the key to its dissociation \cite{5}. Therefore, as for the catalytic behavior of \ce{Fe} surface in  dissociation, the breakdown of the triple bond of the molecule due to the interaction between orbitals of \ce{Fe} surface atoms and the antibonding molecular orbitals needs to be investigated in detail. This paper presents some results of $ab~ initio$ calculations regarding the adsorption and dissociation of nitrogen molecule on \ce{Fe} surface and provides some microscopic details on its dissociation.

\section{2.	Computational Details}

DFT calculations are performed using the Spanish Initiative for Electronic Simulations with Thousands of Atoms (SIESTA)\cite{8} method. Double-$\zeta$ valence orbitals plus single-$\zeta$ polarization $d$ orbitals (DZP)\cite{8} and the Perdew-Burke-Ernzerhof (PBE)\cite{9} generalized-gradient approximation (GGA) functional are used as basis sets and the exchange-correlation energy functional, respectively. Troullier-Martins\cite{10} pseudopotentials are used to take into account the core electrons of the atoms. Pseudopotentials are generated by ATOM package included in SIESTA, treating $4{s^1}3{d^7}4{p^0}$$4{f^0}$ orbitals as valence states for \ce{Fe} and $2{s^2}2{p^3}3{d^0}4{f^0}$ as valence states for \ce{N}. $200~Ry$ cutoff for real space mesh is used and integration over Brillouin zone is done on the $3 \times 3 \times 1$  k-point grid within Monkhorst-Pack scheme\cite{3}. The \ce{Fe}(111) surface is modeled by a five-layer slab of atoms with a vacuum region of 20\AA~above it. A   supercell, with four atoms per monolayer, is adopted for \ce{N2} adsorption simulation. The nitrogen molecule is placed on the top side (in the vacuum region) of the slab. The bottom two atomic layers of \ce{Fe} substrate are fixed and others including two \ce{N} atoms are free to move during the structural relaxation. The atoms are relaxed until the forces on the ions are less than 0.02~eV/\AA. The atomic coordinates are optimized by the conjugate gradient method (CG).

\section{3.	Results and Discussion}
\begin{figure}
\centering
\includegraphics[width=1.0\textwidth]{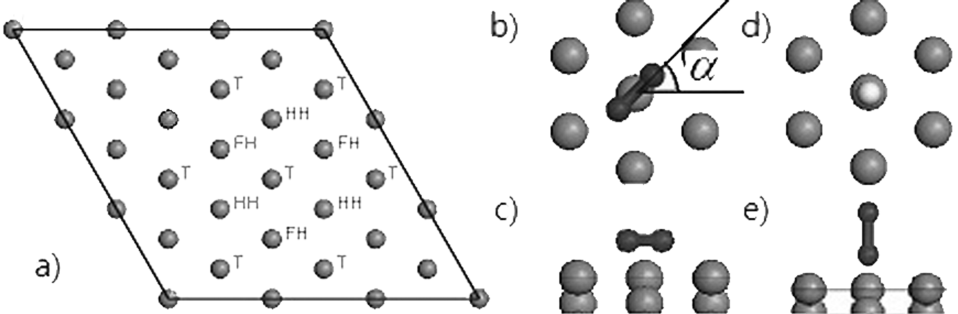}
\caption{a) High symmetry sites on \ce{Fe} (111) surface b), c) Parallel placement of \ce{N2} over a T site with an azimuthal angle of $\alpha$, d), e) Vertical placement of \ce{N2} over a T site} \label{fig:1}
\end{figure}

Before simulating the nitrogen molecule adsorption, \ce{Fe} slab and \ce{N2} molecule are optimized respectively within a supercell of the same size for the consistency of the calculation results. The optimized bond length of \ce{N2} is 1.1091~\AA. Iron exists in different phases at different temperatures. For the simulation, $\alpha$-phase is adopted because under 910~\textcelsius~ it is in $\alpha$-phase with bcc structure (lattice constant 2.86~\AA). Prior to the relaxation, the interlayer distance of the slab is 0.82~\AA. After the relaxation, the distance between the first and second layers is expanded to 0.821~\AA, the distance between the second and third layers is reduced to 0.167~\AA~and the distance between the third and fourth layers is again expanded to 1.004~\AA, which is due to the electron density difference between the outermost layer and inner layers.
To investigate the behavior of the nitrogen molecule, three high symmetry sites on \ce{Fe}(111) surface -? hcp hollow site, fcc hollow site and top site which are denoted as HH, FH and T, respectively, are considered (FIG  \ref{fig:1}).
 
As for \ce{Fe}(111) surface, HH sites are on the third layer from the top, FH sites are on the second layer and T sites are on the topmost layer. As an example, FIG \ref{fig:1} shows two kinds of initial orientation of the nitrogen molecule above the surface, one for a parallel placement, another for a vertical placement, over a T site. To find out dissociation sites on the surface, nitrogen molecules are placed and relaxed over three different symmetry sites mentioned above with various orientations. For a parallel placement, the distance between the center of mass of the nitrogen molecule and the surface is varied from 1.0~\AA~to 3.0~\AA~as well as the azimuthal angle $\alpha$ of the molecule from $-30^{\circ}$ to $+30^{\circ}$. The relaxation shows that only the cases in which molecules are placed within a distance of 1.0~\AA~above an hcp hollow site with an azimuthal angle between $-10^{\circ}$ and $+10^{\circ}$ results in nitrogen dissociation while other cases do not. Since this paper is mainly focused on how the nitrogen molecule is dissociated, here is considered in detail a dissociation case labeled as HH1 in which the molecule is placed parallel to the surface at a distance of 1.0~\AA~from the HH site with  an azimuthal angle of $0^{\circ}$. In addition, here are also considered two cases labeled as HH2, HH3 in which the molecule is aligned parallel to the surface at a distance of 2.0~\AA~and 3.0~\AA~from the surface, respectively, with the same azimuthal angle as in the case HH1. The cases HH2 and HH3 do not result in dissociation, but are very similar to HH1.

\begin{figure}
\centering
\includegraphics[width=1.0\textwidth]{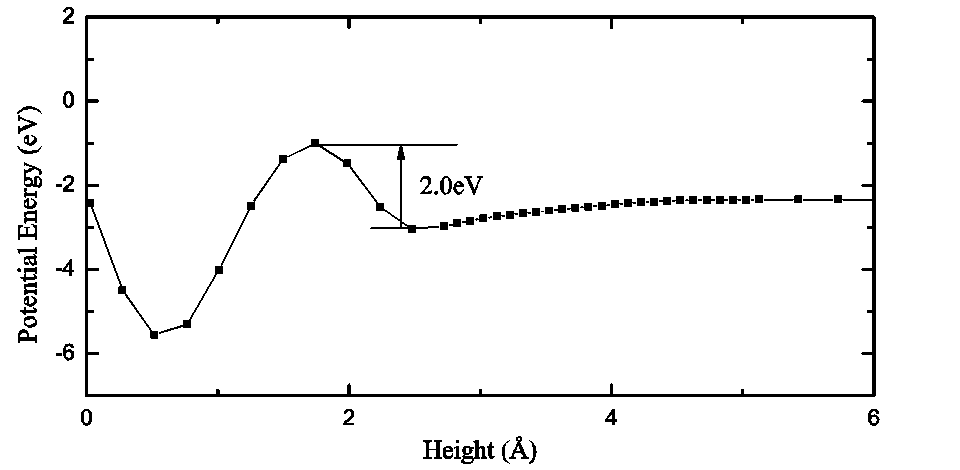}
\caption{Potential energy of \ce{N2} with parallel orientation with an  azimuthal angle of $0^{\circ}$ over an HH site on \ce{Fe}(111) surface} \label{fig:2}
\end{figure}

In FIG \ref{fig:2}, the potential energy of the nitrogen molecule over an HH site is plotted as the distance from the center of mass of the molecule to the surface is varied from 6~\AA~to 0~\AA~. Certainly, the potential energy of a molecule over a surface is very complicated. The curve in FIG \ref{fig:2} is for the extreme case when the molecule is kept parallel to the surface while approaching towards it and maintaining its azimuthal angle at zero. The curve is obtained by calculating the potential energy at each position of the molecule by subtracting the total energy of \ce{Fe} slab and the gas phase \ce{N2} from that of the whole system and interpolating these discrete values. Part of the curve within the range of 1.7~\AA~and 6.0~\AA~represents the chemisorption process and the rest represents the molecular dissociation process. The local maximum around 1.7~\AA~represents the energy barrier for the dissociation of the chemically adsorbed nitrogen molecule and the small barrier near 5.0~\AA~is the energy for the molecular chemisorption. It is clear from FIG \ref{fig:2} that molecules can be dissociated when placed within a distance of 1.7~\AA~from the surface. When it is dissociated, the bond length is about 2.701~\AA. Moreover, it can be seen that the molecule cannot be dissociated but adsorbed on the surface when it is placed at 2~\AA~and 3~\AA~from the surface, respectively, which is consistent with the simulation result that only in the case HH1 the molecule is dissociated. According to the plot, the energy barrier for \ce{N2} dissociation is about 2.0~eV, which is much smaller than 9.8~eV\cite{5} - the dissociation energy of a gas phase nitrogen molecule. This implies that \ce{Fe}(111) surface can be used for catalytic dissociation of molecular nitrogen. Again, it must be kept in mind that such a parallel approach of the molecule towards the surface is an extreme situation in reality.

\begin{figure}
\centering
\includegraphics[width=1.0\textwidth]{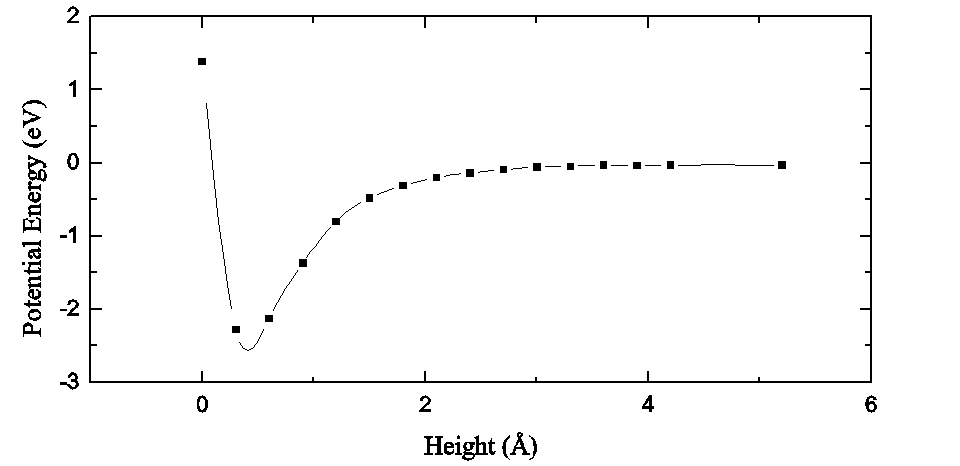}
\caption{Potential energy of vertically oriented \ce{N2} molecule over a T site on \ce{Fe}(111) surface} \label{fig:3}
\end{figure}

Simulation results show that molecules oriented vertically are chemically adsorbed without being dissociated. As an example, in FIG \ref{fig:3}, the potential energy is plotted against the height of the vertically placed nitrogen molecule.
 
In FIG \ref{fig:3}, the local minimum around 0.4~\AA~is not for the dissociated state but for the chemisorped state. The rapid increase in energy around 0~\AA~implies that it is very difficult for the molecule to be dissociated when it approaches towards the surface with vertical orientation. The vertically oriented molecule has to turn via the interaction with the surface before it is aligned parallel to and adsorbed onto the surface. However, in simulations, when it became parallel, the distance between the molecule and the surface was so large (greater than 2~\AA) that it could not be dissociated.

\begin{figure}
\centering
\includegraphics[width=1.0\textwidth]{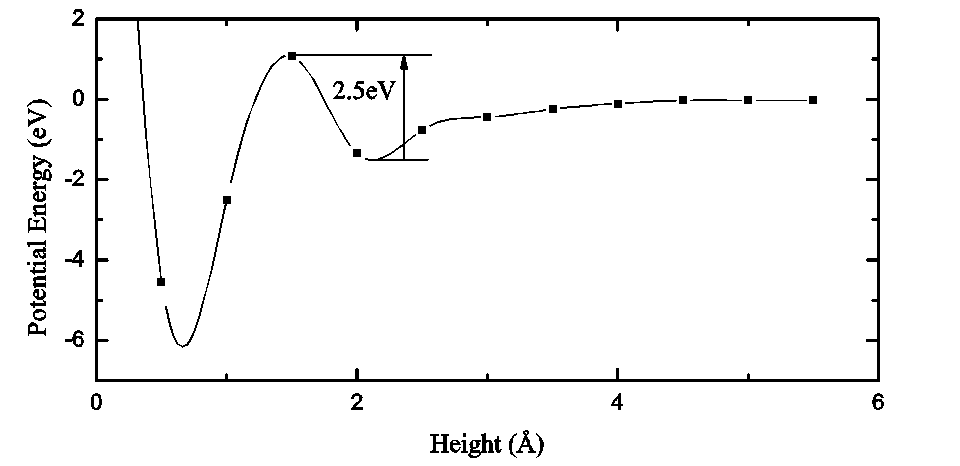}
\caption{Potential energy of \ce{N2} with parallel orientation over \ce{Fe}(100) surface} \label{fig:4}
\end{figure}

Due to its morphology, \ce{Fe}(111) surface is more active in dissociating nitrogen molecules than other surfaces such as \ce{Fe}(100). It has hollow sites that are wider and deeper than the ones on \ce{Fe}(100) surface. Such geometry may provide a favorable condition for the adsorption and dissociation of nitrogen molecule. In fact, the potential energy of the molecule over an \ce{Fe} atom on the topmost layer of \ce{Fe}(100) surface is plotted against the height of the molecule measured from the surface in FIG \ref{fig:4}. Part of the curve within the range from 1.5~\AA~to 5.5~\AA~represents the chemisorption process and the rest is for the dissociation process. It shows that the molecule needs to overcome the energy barrier of 2.5~eV for the dissociation, which is greater than 2.0~eV for \ce{Fe}(111) case. In this case, the bond length is about 3.45~\AA~that is greater than that for \ce{Fe}(111) case.
 
As mentioned above, the activation energy for \ce{N2} dissociation is lowered via the interaction with the surface. To analyze the above result based on the electronic structure, the density of states projected onto $2p$ orbitals of the nitrogen molecule and $3d$ orbitals of \ce{Fe} atoms on the topmost layer is investigated.
 
\begin{figure}
\centering
\includegraphics[width=1.0\textwidth]{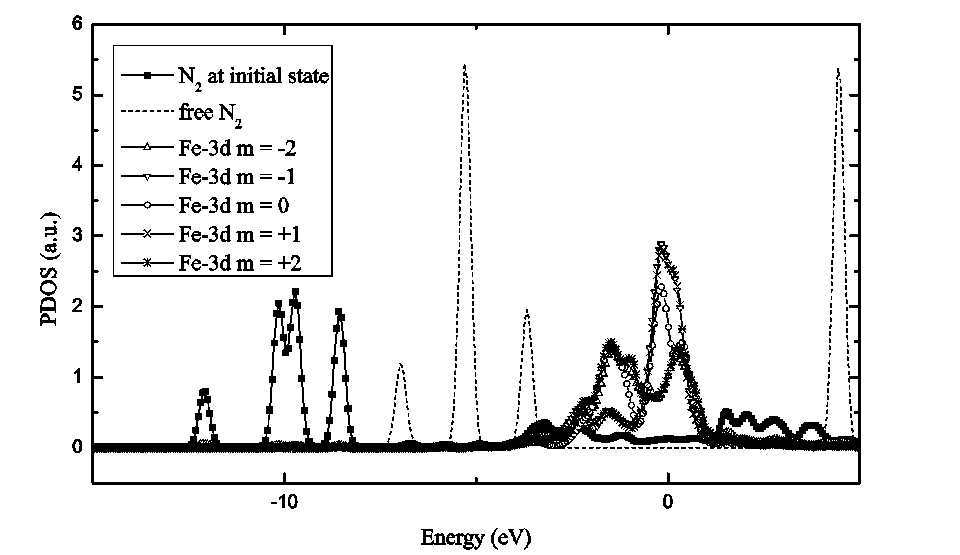}
\caption{PDOS of the gas phase molecule and the whole system in the initial state (case HH1-I) } \label{fig:5}
\end{figure} 

FIG \ref{fig:5} shows the projected density of states (PDOS) onto $2p$ orbitals of the gas phase molecule and the molecule positioned at the height of 1.0~\AA~from the surface. The line labeled as \ce{Fe}~-~3d $m=-2$, for example, represents the PDOS onto $3d$ orbitals of \ce{Fe} atoms on the topmost layer with the magnetic quantum number $m=-2$ when the whole system is put in the initial state. All the PDOS graphs are obtained by setting the Fermi energy equal to zero. The dotted line is the PDOS of the nitrogen molecule in the gas phase. 

During the formation of the nitrogen molecule, $2p$ orbitals of two nitrogen atoms are combined to produce 6 molecular orbitals (MO) $\sigma _x^ * $,  ${\sigma _x}$, $\pi _y^ * $, ${\pi _y}$, $\pi _z^ * $, ${\pi _z}$ among which ${\pi _y}$ and ${\pi _z}$, $\pi _y^ * $  and $\pi _z^ * $ are degenerated so that the energy levels of MOs formed by $2p$ orbitals are split into four levels, which explains the four peaks in the PDOS of the gas phase nitrogen molecule onto 2p orbitals. These peaks corresponds to  ${\sigma _x}$, ${\pi _y}$  and  ${\pi _z}$, $\pi _y^ * $ and $\pi _z^ * $, $\sigma _x^ * $, respectively.
 
\begin{figure}
\centering
\includegraphics[width=1.0\textwidth]{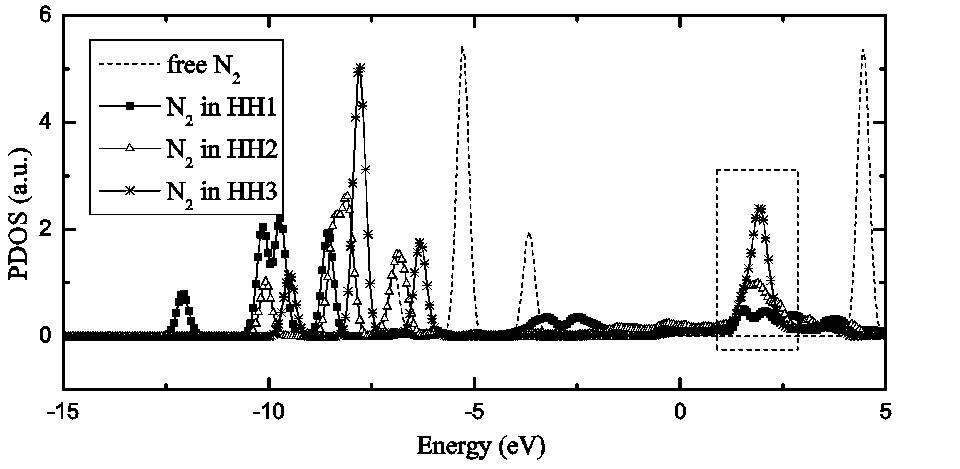}
\caption{MOs of \ce{N2} in cases HH1-I, HH2-I and HH3-I. The dotted line is for MOs of the gas phase nitrogen molecule. The dotted rectangle shows that the closer the molecule approaches to the surface, the broader the highest MO becomes.} \label{fig:6}
\end{figure}

According to the PDOS of \ce{N2} in the state HH1-I (the letter I  means the initial state before its adsorption and dissociation.), among the MOs, only the energetically highest antibonding MO $\sigma _x^ * $  is affected. The lowest three peaks of the PDOS of the gas phase molecule are displaced to the left by about 5~eV when the molecule is put into the initial state, so that $\sigma _x^ * $  of the gas phase molecule is moved and broadened near $3d$ orbitals of \ce{Fe} atoms on the surface. Therefore, it is possible to say that this significant broadening of $\sigma _x^ * $  orbital is due to the interaction with $3d$ orbitals of \ce{Fe} atoms. As for the width of  $\sigma _x^ * $ orbital, the closer the molecule is to the surface, the broader  $\sigma _x^ * $ orbital becomes, and again the stronger the interaction between  $\sigma _x^ * $ and $3d$ orbitals might be.(FIG \ref{fig:6})

Because the antibonding MOs of the molecule need to be filled for the molecular dissociation\cite{5}, the electron population of the nitrogen molecule for each case is investigated. Due to the stronger electronegativity of \ce{N} over \ce{Fe}, the electron density of the \ce{Fe} surface is partially transferred to the molecule. In fact, the electron populations of the gas phase molecule and the molecules in the cases HH1-I, HH2-I, HH3-I are 14.000, 14.300, 14.034 and 13.952, respectively. Interestingly, the electron population of the molecule is the greatest in the dissociation case HH1-I. Moreover, electron populations of $2p$ orbitals of the nitrogen molecule are 7.514, 7.02 and 6.79 for the cases HH1-I, HH2-I, HH3-I, which again shows the greatest value for the dissociation case HH1-I. Considering the broadening of the antibonding MO $\sigma _x^ * $  and the electron population of $2p$ orbitals of the molecule in each case, it can be concluded that the electron density transfer from $3d$ orbitals of \ce{Fe} surface atoms to the antibonding MO $\sigma _x^ * $ may promote the dissociation of nitrogen molecule by filling the MO. However, it is uncertain whether $3d$ orbitals are dominant in electron transfer to the antibonding MO $\sigma _x^ * $.

\section{4.	Summary}

In this paper, the adsorption and dissociation of nitrogen molecule on \ce{Fe}(111) surface is discussed in accordance with the potential energy, the PDOS and the electron population of the nitrogen molecule. The molecule is dissociated on the \ce{Fe}(111) surface when it is placed parallel to the surface with an azimuthal angle between $-10^{\circ}$ and $+10^{\circ}$. The vertically oriented molecule needs to rotate to acquire a parallel position before it can be adsorbed or dissociated on the surface. Therefore, the parallel orientation is favorable in the molecular dissociation of nitrogen on the surface. Moreover, Fe(111) surface significantly lowers the energy barrier for \ce{N2} dissociation. In the dissociation of nitrogen molecule, the interaction between antibonding molecular orbitals of \ce{N2} and $3d$ orbitals of \ce{Fe} atoms on the surface may be a promoting factor: it may allow electron transfer from $3d$ orbitals to the antibonding MOs to lower the energy barrier for  and accelerate the dissociation of nitrogen molecule.

\end{document}